\documentclass[12pt,manuscript]{emulateapj}

\usepackage{url}
\usepackage{natbib}
\usepackage{amsmath}

\newcommand{\beq}{\begin{equation}}
\newcommand{\eeq}{\end{equation}}
\newcommand{\bdi}{\begin{displaymath}}
\newcommand{\edi}{\end{displaymath}}
\newcommand{\IRAS}{\textit{IRAS}}

\newcommand{\arc}{${\rm arcmin}^{-1}$}

\newcommand{\degree}{$^{\circ}$}

\newcommand{\betaval}{2}
\newcommand{\apt}{$16.9 \pm 0.7$} 
\newcommand{\act}{$17.3 \pm 0.3$}
\newcommand{\cpt}{$19.9 \pm 1.3$}
\newcommand{\cct}{$18.6 \pm 0.5$}

\newcommand{\camp}{$(1.60 \pm 0.05) \times 10^{-3}$} 
\newcommand{\aamp}{$(3.03 \pm 0.15) \times 10^{-3}$}
\newcommand{\cslope}{$-2.60 \pm 0.07$}
\newcommand{\aslope}{$-3.14 \pm 0.10 $}

\newcommand{\cirisamp}{$1.61\pm 0.22$} 
\newcommand{\airisamp}{$0.63 \pm 0.03$} 
\newcommand{\cirislope}{$-2.71 \pm 0.07$}
\newcommand{\airislope}{$-3.05 \pm 0.02$} 

\newcommand{\ciav}{260} 
\newcommand{\aiav}{120} 

\newcommand{\cmarcamp}{4.3} 
\newcommand{\cmarcslope}{$-3.4$}
\newcommand{\amarcamp}{0.59}
\newcommand{\amarcslope}{$-3.3$}

\newcommand{\cl}{79.89}
\newcommand{\cb}{0.47}
\newcommand{\al}{45.85}
\newcommand{\ab}{-0.12}

\slugcomment{Submitted to the Astrophysical Journal}

\begin{document}
\shorttitle{Cirrus Power Spectra}
\shortauthors{Roy, A.~et al.}

\title{BLAST05: Power Spectra of Bright Galactic Cirrus at
  Submillimeter Wavelengths}


\author{Arabindo~Roy,\altaffilmark{1}
	Peter~A.~R.~Ade,\altaffilmark{2}
        James~J.~Bock,\altaffilmark{3,4}
	Edward~L.~Chapin,\altaffilmark{5}
	Mark~J.~Devlin,\altaffilmark{6}
	Simon~R.~Dicker,\altaffilmark{6}
	Matthew~Griffin,\altaffilmark{2}
	Joshua~O.~Gundersen,\altaffilmark{7}
        Mark~Halpern,\altaffilmark{5}
        Peter~C.~Hargrave,\altaffilmark{2}
	David~H.~Hughes,\altaffilmark{8}
	Jeff~Klein,\altaffilmark{6}
	Gaelen~Marsden,\altaffilmark{5}
        Peter~G.~Martin,\altaffilmark{1,9}
        Philip~Mauskopf,\altaffilmark{2}
	Marc-Antoine~Miville-Desch{\^e}nes,\altaffilmark{10}
	Calvin~B.~Netterfield,\altaffilmark{1,11}
        Luca~Olmi,\altaffilmark{12,13}
        Enzo~Pascale,\altaffilmark{2}
	Guillaume~Patanchon,\altaffilmark{14}
	Marie~Rex,\altaffilmark{6}
        Douglas~Scott,\altaffilmark{5}
	Christopher~Semisch,\altaffilmark{6}
	Matthew~D.~P.~Truch,\altaffilmark{15}
	Carole~Tucker,\altaffilmark{2}
        Gregory~S.~Tucker,\altaffilmark{15}
	Marco~P.~Viero,\altaffilmark{1}
	Donald~V.~Wiebe\altaffilmark{5}}

\altaffiltext{1}{Department of Astronomy \& Astrophysics, University of Toronto, 50 St. George Street, Toronto, ON  M5S~3H4, Canada}

\altaffiltext{2}{Department of Physics \& Astronomy, Cardiff University, 5 The Parade, Cardiff, CF24~3AA, UK}

\altaffiltext{3}{Jet Propulsion Laboratory, Pasadena, CA 91109-8099}

\altaffiltext{4}{Observational Cosmology, MS 59-33, California Institute of Technology, Pasadena, CA 91125}

\altaffiltext{5}{Department of Physics \& Astronomy, University of British Columbia, 6224 Agricultural Road, Vancouver, BC V6T~1Z1,Canada}
\altaffiltext{6}{Department of Physics and Astronomy, University of Pennsylvania, 209 South 33rd Street, Philadelphia, PA 19104}

\altaffiltext{7}{Department of Physics, University of Miami, 1320 Campo Sano Drive, Carol Gables, FL 33146}

\altaffiltext{8}{Instituto Nacional de Astrof{\'i}sica {\'O}ptica y Electr{\'o}nica (INAOE), Aptdo. Postal 51 y 72000 Puebla, Mexico}

\altaffiltext{9}{Canadian Institute for Theoretical Astrophysics, University of Toronto, 60 St. George Street, Toronto, ON M5S~3H8, Canada}

\altaffiltext{10}{Institut d' Astrophysique Spatiale, UMR8617, Universit{\'e} Paris-Sud, F-91405, Orsay, France}

\altaffiltext{11}{Department of Physics, University of Toronto, 60 St. George Street, Toronto, ON M5S~1A7, Canada}

\altaffiltext{12}{Istituto di Radioastronomia, Largo E. Fermi 5, I-50125, Firenze, Italy}

\altaffiltext{13}{University of Puerto Rico, Rio Piedras Campus, Physics Dept., Box 23343, UPR station, San Juan, Puerto Rico}

\altaffiltext{14}{Laboratoire APC, 10, rue Alice Domon et L{\'e}onie Duquet 75205 Paris, France}

\altaffiltext{15}{Department of Physics, Brown University, 182 Hope Street, Providence, RI 02912}

\begin{abstract}
We report multi-wavelength power spectra of diffuse Galactic dust
emission from BLAST observations at 250, 350, and 500~\micron\ in
Galactic Plane fields in Cygnus~X and Aquila. These submillimeter
power spectra statistically quantify the self-similar structure
observable over a broad range of scales and can be used to assess the
cirrus noise which limits the detection of faint point sources.  The
advent of submillimeter surveys with the \textit{Herschel Space
  Observatory} makes the wavelength dependence a matter of interest.
We show that the observed relative amplitudes of the power spectra can
be related through a spectral energy distribution (SED).  Fitting a
simple modified black body to this SED, we find the dust temperature
in Cygnus~X to be \cpt\ K and in the Aquila region \apt\ K. Our
empirical estimates provide important new insight into the substantial
cirrus noise that will be encountered in forthcoming observations.
\end{abstract}



\section{Introduction} \label{sec:intro}
Energy is being injected continually into the interstellar medium
(ISM) through spiral shocks, violent outflows from massive protostars,
stellar winds, expanding \ion{H}{2} regions and supernova explosions. The
result is a turbulent medium where the dust is well mixed in the
structures produced.  The emission from the relatively nearby ISM at
high Galactic latitude is known as Galactic Cirrus.  Cirrus-like
structure in the brighter emission near the Galactic Plane has been
called `interstellar froth' \citep{waller1994}; this might be an
interesting distinction with a physical basis, but here we will simply
use the term `cirrus'.  The dynamics of the ISM appear to have made
the distribution of density structures self-similar, with fluctuations
in column density and surface brightness present on all observable
scales, though decreasing towards smaller scales.  Statistical
description of random fluctuations through structure functions is an
important method of extracting physical properties hidden in diffuse
emission as well as for providing a quantitative measure to compare
with simulations.
A common statistical tool used to estimate the level of cirrus noise
is the power spectrum.  This is valid for Gaussian random fields.
However, \citet{gautier} found evidence for non-Gaussianity and recent
studies by \citet{ma-dust} have revealed non-vanishing skewness and
excess kurtosis in the underlying brightness fluctuation fields.
Nevertheless, for estimating the variance the power spectrum is still
indicative.

Using power spectra, \citet{gautier} quantified how fluctuations
associated with Galactic cirrus are a source of confusion noise which
limits the detectability of point sources. Even with the improved
angular resolution of imagers like SPIRE and PACS on the
\textit{Herschel Space
  Observatory}\footnote{http://www.esa.int/SPECIALS/Herschel/index.html}
(hereafter \textit{Herschel}), this `cirrus noise' remains important,
often dominant. Therefore, to assess the noise it is vital to know the
statistical properties of the interstellar diffuse emission at the
relevant wavelengths.

A number of other statistical analyses of emission have been
carried out on the basis of power spectra for high latitude clouds,
e.g., by \citet{kiss-galcir}, \citet{jeong2005}, and \citet{ma-dust}.
In this paper we analyse diffuse dust emission in the Galactic Plane
and present for the first time multi-wavelength power spectra in the
submillimeter, based on observations in Cygnus~X (Cyg~X) and Aquila
with BLAST \citep{pascale2008} at 250, 350, and 500~\micron.
We also analyse the same regions at 100 and 60~\micron\ using IRIS,
\IRAS\ data reprocessed by \citet{mairis}.
Compared to high latitude studies, analysing diffuse emission in the
Galactic Plane in terms of the structure of the ISM resulting from
its turbulent properties is more challenging because of the long path
lengths, high column density, high star-formation rate, and
contamination by compact sources. In addition, in the Galactic Plane, 
self-gravity can play an important role in shaping diffuse structures
at smaller scales.
Of particular importance are the potentially strong spatial variations
of the radiation field due to star formation activity.  As discussed
further in \S~\ref{subsec:masscol}, this could produce variations in
the dust emission independently of any changes in structure of the
ISM.

Our paper is organized as follows.
We begin with a brief description of BLAST observations
(\S~\ref{sec:obs}) and then introduce important aspects of the power
spectrum and accompanying cirrus noise (\S~\ref{sec:ps}).
We analyse IRIS data in \S~\ref{sec:resI}, placing this in the context
of earlier studies and providing a short-wavelength reference for our
submillimeter studies.
In \S~\ref{sec:resB} we examine the BLAST data: the noise; effect of
the point spread function (PSF) or beam; removal of compact sources;
and the exponents and amplitudes of the submillimeter power spectra.
We estimate the cirrus noise for these maps and compare it with the
completeness depth of the BLAST05 point-source catalogs at
250~\micron. In \S~\ref{sec:depth} we also discuss some implications
for related approved observations with \textit{Herschel}.
We show in \S~\ref{sec:dis} how the observed wavelength dependence of
the amplitude of the power spectrum can be understood as a
straightforward consequence of the spectral energy distribution (SED)
of the dust and we fit a simple modified black body to estimate the
dust temperature.
Our empirical results provide new insight into what cirrus noise might
be expected in submillimeter observations.

\section{Observations} \label{sec:obs}
   
In 2005, the Balloon-borne Large Aperture Submillimeter Telescope
(BLAST05; \citealp{pascale2008}) made unbiased surveys in targeted
regions of the Galactic Plane.  \citet{chapin2008}  analysed 
a 4~deg$^2$ field in Vulpecula and here we used data from the 
two largest surveys. Aquila (Rivera-Ingraham et al., in preparation) 
is a  6~deg$^2$ region observed for 6.1 h, while Cyg~X (Roy et al., in
preparation) covers 10~deg$^2$ observed over 10.6 h. Both have
good cross-linking from orthogonal scanning. The maps were made with
the SANEPIC algorithm \citep{patanchon2008} and were calibrated using the procedure
discussed in \citet{truch2008}.

Although a 2-m telescope, BLAST05 produced maps of only $3$\arcmin\ resolution (see
\S~\ref{sec:beam}) due to an anomalous PSF, corrupted by some
uncharacterized combination of mirror distortion and de-focus
\citep{truch2008}.\footnote{This problem was fixed for the 2006 flight from
  Antartica  \citep{truch2009}.} Nevertheless, maps from the 2005
  flight have high signal-to-noise and are oversampled with
  15\arcsec\ pixels, so that Lucy-Richardson (L-R) deconvolution can
  be used to improve the resolution significantly (Roy et al., in
  preparation). This goal is particularly important for extracting
  point sources (\S~\ref{sec:sr}) but otherwise not essential for the
  study of diffuse emission.
We analysed the two largest surveys, selecting in Aquila a square
sub-field of size $1.83$\degree\ centred on $l = \al$\degree , $b =
\ab$\degree\ and a similar-sized field in Cyg~X centred on $l = \cl
$\degree , $b = \cb$\degree .
The latter does not include the brighter star-forming regions to the
east containing W75-N and DR21 \citep{schneider2006}.

We also analysed IRIS maps at 100 and 60~\micron\ toward these
selected regions. These maps have $\sim 4$\arcmin\ resolution on
1.5\arcmin\ pixels \citep{ma-iras} and we used the version in which
the sources have been removed by the technique described by
\citet{ma-lagache}.

\section{Power Spectrum and Cirrus Noise}\label{sec:ps}

The power spectrum is the Fourier transform of the auto-correlation
function of the intensity map $I(x,y)$.  In $k$-space, $P(k_x,k_y)$ is
simply related to an image by
\begin{equation}
P(k_x,k_y)=\langle \tilde{I}(k_x,k_y)\tilde{I}^\star(k_x,k_y) \rangle,
\label{eq:pseck}
\end{equation}
where $\tilde{I(}k_x,k_y)$ is the Fourier transform of the image and
$\tilde{I}^\star$ its complex conjugate. We used the IDL routine FFT
to compute the two-dimensional Fourier transforms.  To ensure a smooth
periodic boundary condition near the edges \citep{ma-iras}, the maps
were apodized by a sine function over a range 10\% the width of the
map.
The power spectrum $P(k)$ is obtained by averaging $P(k_x,k_y)$ over
an annulus placed at $k = \sqrt{k^2_x+k^2_y}$.

In practice, contributions to the total power spectrum come not only
from diffuse dust emission, but also point sources, the cosmic
infrared background (CIB), and the noise.  When these components are
statistically uncorrelated, the total power spectrum can be expressed
as \citep{ma-dust}
\begin{equation}
P(k)=\Gamma(k)\left[P_{\rm{cirrus}}(k)+P_{\rm{source}}(k)+P_{\rm{CIB}}(k)\right]+N(k),
\end{equation}
where $\Gamma(k)$ is the power spectrum of the PSF (the square of the
modulus of the two-dimensional Fourier transform of the PSF), which
decays at large $k$.  For the bright Galactic Plane fields targeted
here, the contribution from the CIB to the power spectrum is
negligible. While the noise is measurable (\S~\ref{pow250}), it too
makes an insignificant contribution.

\citet{gautier}, followed by many other authors
\citep{kiss2001,kiss-galcir,ma-iras}, have shown that the power
spectrum of Galactic cirrus follows a power law
\begin{eqnarray}
P(k)=P(k_0)(k/k_0)^{\alpha},
\label{pow}
\end{eqnarray}
quantified by an amplitude $P_0\equiv P(k_0)$ at some fiducial $k_0$ and an
exponent $\alpha$ that is typically $-3$ \citep{ma-dust}.
From their analysis, the associated `cirrus noise' for a telescope
with mirror diameter $D$ working at wavelength $\lambda$ can be
quantified as
\begin{equation}
\begin{split}
\sigma_{\rm{cirrus}} = 100\ (r/1.6)^{2.5} \left(\frac{\lambda/250\ {\rm
    \micron}}{D/3.5\ {\rm m}}\right)^{2.5} \\
    [P(k=0.1\ {\rm
    arcmin}^{-1})/10^{-3}\ {\rm MJy}^2\ {\rm sr}^{-1}]^{0.5}\ {\rm
  mJy}.
\label{sigcir}
\end{split}
\end{equation}
Here our assumptions about the measurement strategy for point sources
are the same as made by \citet{helou1990} and adopted by
\citet{kiss2001}, which in the notation of \citet{gautier} are a
measuring aperture $d$, with `resolution ratio' $r = d/(\lambda/D)$ of $1.6$ 
and a reference
annulus with `separation ratio' 2, and also $\alpha = -3$, close to
what we find below. The beam-related factor quantifies the effect of a
smaller beam probing smaller spatial scales where the power in the
fluctuations is weaker.

If the sources being measured are extended, as will be the case in
many Galactic surveys, then the measuring aperture needs to be larger.
The main consequence of this on increasing $\sigma_{\rm{cirrus}}$ is
captured by the factor $r^{2.5}$ for the range of interest (see also
Fig.~3 in \citealp{gautier}).

Working with 100~\micron\ \IRAS\ data on fields of different average
surface brightness $\langle I_{100} \rangle$, and adopting a fiducial
scale $k_0 = 0.01 $ \arc\ at which the amplitude is $P_{100}$,
\citet{gautier} found the trend that $P_{100} = C \langle I_{100}
\rangle^3$, where $C$ is a proportionality constant.  If this is
substituted in the above, we recover the formula given by
\citet{helou1990} and evaluated by \citet{kiss2001} estimating the 
cirrus noise for different levels of cirrus brightness.
Note that the cirrus noise estimate described by \citet{ma-dust} used
a slightly different definition of the noise, as well as 
incorporating a trend giving a slightly steeper power law with
increasing $\langle I_{100} \rangle$, and in effect a lower $C$, which
together conspire to lower the noise estimate by about a factor six
for bright cirrus.

For reasons discussed below, both estimating $P_{100}$ from $\langle
I_{100} \rangle$ and scaling it to the appropriate wavelength of
observation are problematical. Therefore, if possible the exponent and
amplitude of $P(k)$ should be measured directly for the field of
interest at the relevant wavelength.  Measuring and normalizing at a
scale as close as possible to the beam also avoids issues of
extrapolation.  With \textit{Herschel} in mind, we have chosen $k =
0.1$ \arc.  The normalization of the amplitude in equation
(\ref{sigcir}) anticipates what we find at 250~\micron\ for the two
bright fields we examined.

\section{IRIS Power Spectra}\label{sec:resI}

Our primary goals are to analyse statistical fluctuations of diffuse
dust emission in the Galactic Plane using the power spectrum, and to
measure the wavelength dependence of the amplitude of this power
spectrum.  To connect with previous work and provide a reference at
shorter wavelengths, we begin with IRIS.
IRIS (like \IRAS) data come in `plates' 12.5\degree\ on a side.  We
computed the power spectrum for the entire plate containing the Aquila
field (plate 263) but because of a gap in the \IRAS\ sky coverage near
Cyg~X (plate 361) we had to settle for a smaller region, 6\degree\ on
a side in Galactic coordinates centred on $l = 80.69 $\degree\ and $b
= 0.76$\degree . Binned estimates of $P(k)$ and their uncertainties
were evaluated as in \S~\ref{2dps}.

\begin{figure}
\includegraphics[angle=0,width=\linewidth]{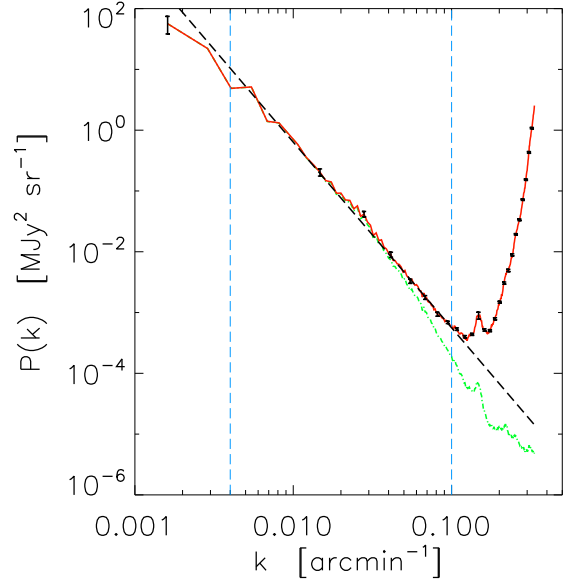}
\caption{IRIS 100~\micron\ power spectrum (dot-dash curve) of a square
  12.5\degree\ on a side in Aquila spanning the Galactic Plane; and
  after dividing by $\Gamma(k)$ (solid).  The error bars, plotted only
  every tenth point for clarity, do not account for the imperfect
  approximation to the beam whose effect is clearly present at large
  $k$.  We fit the power law (dashed line) over the range
  $0.004\ \rm{arcmin^{-1}} <$ $k$ $ < 0.1 \ \rm{arcmin^{-1}}$ marked
  by the vertical dashed lines.  The exponent is \airislope\ and
  $P_{100}=$ \airisamp\ $ {\rm MJy}^2\ {\rm sr^{-1}}$.  }

\label{fig:ps100}
\end{figure}

The power spectrum for the Aquila field at 100~\micron\ is shown in
Figure \ref{fig:ps100}; the power spectrum for the Cyg~X field (not
shown) is similar.  The noise, $N(k)$, can be estimated by comparing
different independent \IRAS\ maps (HCONs; \citealp{ma-iras})and for
these bright fields is negligibly small.  The behavior of $P(k)$ 
is very much like that seen at higher
latitudes and with BLAST (\S~\ref{pow250}), except that here it does
not fall off at the highest $k$ quite as fast as expected from the
effect of the beam \citep{ma-iras}.  We have examined a number of
plates crossing the Galactic Plane as well as the adjacent plates at
higher latitude and conclude that this is an effect seen only in the
Plane.  
Apparently in these source-removed maps there are significantly 
larger fluctuations within the scale of the beam than expected,
and subsequently these are greatly amplified by the beam correction.
These fluctuations are likely an artifact of the source removal 
technique, which would be maximized in the Galactic Plane, where 
the surface density of sources is greatest.  For the beam $\Gamma(k)$ 
we used a Gaussian as specified by \citet{ma-iras} which is clearly
just an approximation to the complicated effective beam of the 
images formed from the IRAS timestreams.  In any case, here we 
use only the part of the power spectrum at lower $k < 0.1$~\arc, 
which is largely unaffected by the beam correction.
We also avoid small values of $k < 0.004$~\arc\ where the power is
significantly affected by the apodization of the image edges
\citep{ma-iras}.  Figure~\ref{fig:ps100} confirms empirically that
this is a good choice for the range of $k$ used for the power law fit.
The non-linear weighted fits were carried out with the IDL routine.

For 100~\micron\ we find an exponent $\alpha$ of \cirislope\ in Cyg~X
and \airislope\ in Aquila.
\cite{ma-dust} find that the steepness of the power spectrum increases
as a slow function of $\langle I_{100} \rangle$.  For the two fields
$\langle I_{100} \rangle$ is \ciav\ and \aiav~${\rm MJy}\ {\rm
  sr^{-1}}$, respectively, and the expected exponents from this trend
are \cmarcslope\ and \amarcslope, slightly steeper than what we find.
Note however that they find a scatter of 0.3 about their trend.
Note also that while we report the formal errors from the fit to the
power spectrum, systematic errors in the derived exponent could be
larger, on the order of 0.1.

We find $P_{100}$ = \cirisamp\ and \airisamp\ ${\rm MJy}^2\ {\rm
  sr^{-1}}$ for Cyg~X and Aquila, respectively.
These can be compared to values of \cmarcamp\ and \amarcamp\ $ {\rm
  MJy}^2\ {\rm sr^{-1}}$ from the trend in \citet{ma-dust}, about
which there is a factor of three scatter at high $\langle I_{100}
\rangle$.  Again the agreement is better for the Aquila field, despite
the fact that there are significant asymmetries in the image (the
bright swath of the Galactic Plane and residual striping) which make
the two-dimensional power spectrum not quite circularly symmetric.

For more direct comparison with the BLAST power spectra, we computed
the IRIS power spectra for the smaller BLAST sub-fields.  The
exponents of the power spectra are $-2.97 \pm 0.23$ and $-2.82 \pm
0.10$, respectively.  These are harder to measure, given the smaller
dynamic range, but do not appear to have changed significantly despite
the larger $\langle I_{100} \rangle$ values ($430$ and $330\ {\rm
  MJy}\ {\rm sr^{-1}}$).
Expressing amplitudes as $P_{100}$ (though $k = 0.01$ \arc\ is beyond
our range in these small sub-fields), we find $3.1 \pm 1.3$ and
$0.75\pm 0.11 \ {\rm MJy}^2\ {\rm sr^{-1}}$, respectively, compared to
trend values of 29 and 13 ${\rm MJy}^2\ {\rm sr^{-1}}$.
Because of the $P^{1/2}$ dependence in equation~(\ref{sigcir}),
predictions of cirrus noise depend less strongly on any deviations
from the trends.  Nevertheless, these results illustrate the important
point that when attempting to assess the cirrus noise, one should, if
possible, measure the exponent and amplitude near the spatial
frequencies of interest.

Because the cirrus noise is wavelength dependent, ideally $P_0$ would
be measured at each relevant wavelength too. The amplitude might in
principle be scaled, say from $P_{100}$.  Not all scaling
prescriptions in the literature can be valid, however, and we discuss
our recommendation in \S~\ref{sec:dis}.

\section{BLAST Power Spectra}\label{sec:resB}

\subsection{Noise Power Spectrum}\label{sec:noise}

\begin{figure}
\includegraphics[width=\linewidth,angle=0]{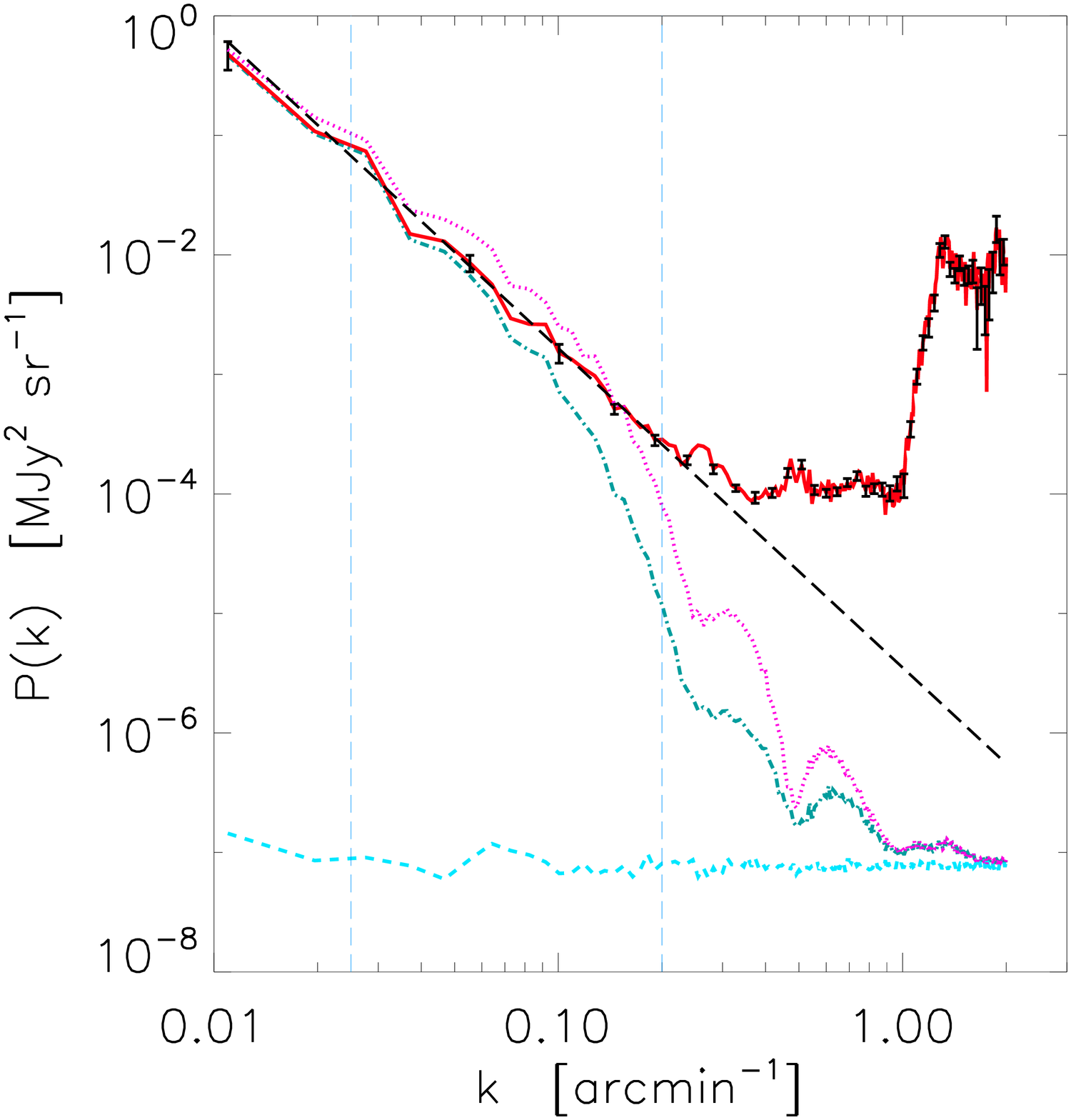}
\caption{BLAST power spectra at 250~\micron\ for the Cyg~X region
  shown in Fig.~\ref{fig:opsc}. The dotted curve, corresponding to the
  upper panel of Fig.~\ref{fig:opsc}, shows that a significant amount
  of power is present at intermediate scales due to compact
  sources. The lower dot-dash line is $P(k)$ of the same region after
  removing sources (corresponding to Fig.~\ref{fig:psthumb}). The
  oscillatory behavior at large $k$ is due to the corrupted BLAST05
  PSF (Fig.~\ref{fig:tf}).  The almost horizontal dashed line is the
  power spectrum of the noise map (\S~\ref{sec:noise}). At very small
  scales where the astronomical signals become correlated within the
  beam, the power spectra meet the noise level. The solid line is the
  power spectrum $P_{\rm{cirrus}}(k)$, after subtracting the noise and
  dividing by $\Gamma(k)$ to remove the effect of the beam (as in
  Fig.~\ref{fig:ps100}, error bars at high $k$ do not include the
  imperfect knowledge of the beam and noise).  The vertical dashed
  lines indicate the restricted range $0.025\ \rm{arcmin^{-1}} <$ $k$
  $ < 0.2\ \rm{arcmin^{-1}} $ used for fitting the power law.  The
  exponent is \cslope\ and $P(0.1\ \rm{arcmin^{-1}})\ =\ $ \camp\ $
  {\rm MJy}^2\ {\rm sr^{-1}}$.}
\label{fig:ps250}
\end{figure}

In the BLAST map-making procedure \citep{patanchon2008}, a variance
map $V(x,y)$ is produced based on the noise in the time-stream data
and the map coverage by the bolometers in the arrays.  For these BLAST
surveys, the resulting maps of $V$ are quite uniform.  A realization
of the noise map $N(x,y)$ can be constructed for each independent,
uncorrelated pixel from $N(x,y)=\sqrt{V(x,y)} v ,$ where $v$ is a
Gaussian random variable with unit standard deviation and zero mean.
Given the relative uniformity of the variance map, the resulting
simulated noise map is close to white on all scales and thus the power
spectrum of this noise is quite flat (see the example for Cyg~X at
250~\micron\ in Fig. \ref{fig:ps250}).  At large $k$ the total power
spectrum decays to this noise level, because of the combined effects
of the decreasing power in the diffuse cirrus emission and the PSF.
There is also a component of low frequency noise in the map arising
from the very long time scale $1/f$ noise present in the time streams.
However, in cross-linked maps produced using data from multiple scanning 
directions this is greatly reduced by the SANEPIC algorithm (e.g., Fig.~10
in \citealp{patanchon2008}) and is not important here because the cirrus
signal at small $k$ is so large.

\subsection{Effect of the Beam}\label{sec:beam}

\begin{figure}
\includegraphics[angle=0,width=\linewidth]{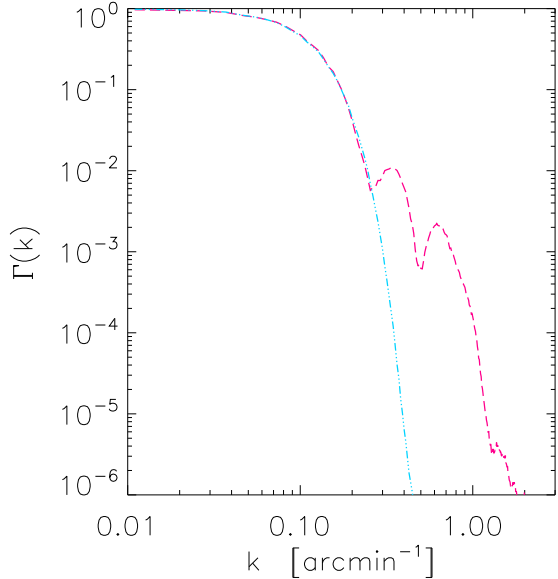}
\caption{$\Gamma(k)$ for BLAST05 in Cyg~X at 250~\micron\ (dashed
  curve).  The dot-dash line shows how this can be approximated by the
  power spectrum of a Gaussian PSF of FWHM 3.3\arcmin.}
\label{fig:tf}
\end{figure}

The BLAST05 PSF in telescope coordinates is shown in
\citet{truch2008}.  A synthetic beam can be made for a particular map,
taking into account the scan angles and coverage \citep{chapin2008}.
Because this synthetic beam is not known out to the full size of the
map, we have not derived its power spectrum $\Gamma (k)$ directly.
Instead, we convolved the noise map with the PSF directly, found the
power spectrum of that map, and divided by the power spectrum of the
noise map.  Figure \ref{fig:tf} shows $\Gamma(k)$ for the BLAST05 PSF
at 250~\micron\ for the Cyg~X field.  This has a number of features
characterizing scales seen in the corrupted PSF.  Clearly, the
observed $P(k)$ will be seriously suppressed at high $k$ and so we
will use data only for $k < 0.2$ \arc.  At these scales, where the
beam correction is not too large and so both more reliable and of less
import, $\Gamma(k)$ can be described by a Gaussian with $\sigma_\Gamma
= 0.08$~\arc.  Using the Fourier relation $2\sqrt 2 \pi
\sigma_\Gamma\sigma_{\rm{b}} = 1$, this corresponds to a Gaussian beam
of $\sigma_{\rm{b}} = 1.4$\arcmin, or FWHM $= 3.3$\arcmin, slightly
smaller than for IRIS. This is close to the 3.1\arcmin\ full width at
half-power found by \cite{truch2008}. The corruption of the BLAST05 PSF is such that
the initial falloff in $\Gamma(k)$ is rather similar for the three
wavelengths, and for both fields.

\subsection{Source Removal} \label{sec:sr}

\begin{figure}
\includegraphics[angle=0,width=\linewidth]{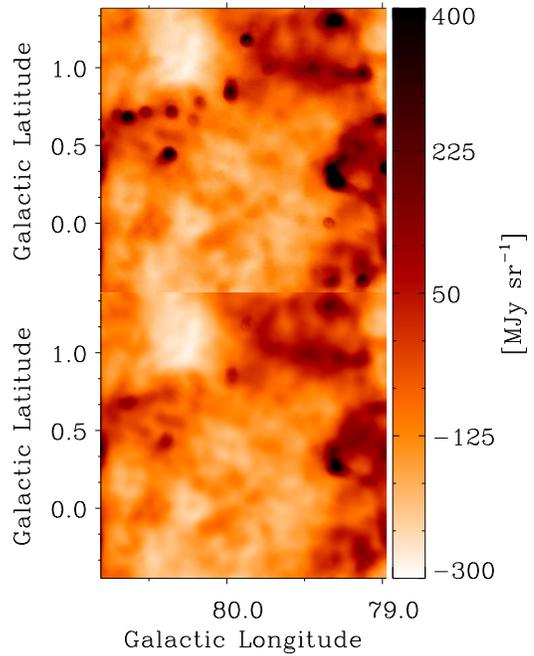}
\caption{BLAST05 map at 250~\micron\ of the selected region of Cyg~X
  used for studying diffuse emission. The map in the upper panel
  includes point sources which, due to the corrupted BLAST05 beam, are
  compact structures of size $\sim 3.3$\arcmin.  In the lower panel
  the sources have been removed.}
\label{fig:opsc}
\end{figure}

In the Galactic Plane, the power spectrum is seriously contaminated by
compact sources. To remove them, we first deconvolve the BLAST maps
using the L-R method which, importantly, conserves flux. Compact
sources are of size $\sim 1.5$\arcmin\ in these maps and more easily
identified. They are fairly well described by Gaussians.
We fit Gaussians (multiple if crowded) to obtain flux densities,
positions, and FWHM (major axes and position angle) of the compact
sources.
We then convolve these Gaussians with the synthetic beam and subtract
them from the original maps.
For the Cyg~X sub-field analysed, the upper and lower panels of
Figure~\ref{fig:opsc} show BLAST maps at 250~\micron\ before and after
source removal, respectively. Faint residuals appear near some of the
brightest sources, because of multiscale structure in the ISM in which
they are embedded.

\subsection{Two-dimensional Power Spectrum} \label{2dps}

\begin{figure}
\includegraphics[angle=0,width=\linewidth]{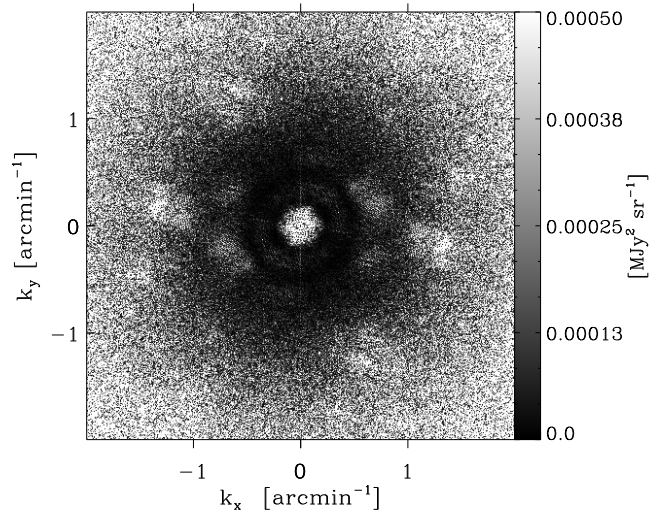}
\caption{Two dimensional power spectrum multiplied by $(k/k_0)^3 $
  (with $k_0\ =\ 0.1$ \arc ), for the Cyg~X map with compact sources
  removed, as shown in the lower panel of Fig.~\ref{fig:opsc}.  Dark
  rings are an imprint from the BLAST05 PSF
  (Fig.~\ref{fig:tf}). Scaling by $k^3$ has amplified noise at large
  $k$ and highlighted the bright central plateau, which contains the
  information most relevant for this study.}
\label{fig:psthumb}
\end{figure}

Figure \ref{fig:psthumb} shows the two-dimensional power spectrum for
the Cyg~X region at 250~\micron. Because $\alpha \sim -3$ produces a
large intrinsic dynamic range in the power spectrum, we have
multiplied it by $(k/k_0)^3$, with $k_0\ =\ 0.1$ \arc.  The dark rings
are produced by the same features giving the dips in $\Gamma(k)$.
Noise is being amplified at large $k$ because of the $k^3$ multiplier.

The two-dimensional power spectrum is circularly symmetric, justifying
the annular averaging to find the mean $P(k)$.  Most of the
information relevant to the diffuse emission is contained in the
plateau in the central region of Figure~\ref{fig:psthumb}, and the
average there in this representation is close to $P(k = 0.1\
\rm{arcmin}^{-1})$.

The error in $P(k)$ for each annulus is the standard error of the
mean.  While removing the effect of beam from the power spectrum, we
have only scaled this error by the inverse beam, not accounting for
the fact that its shape is not perfectly known at high $k$.
 However, this is of no consequence because we do not fit the
data at large $k$ where the beam correction is significant.

\subsection{Exponent and Amplitude}\label{pow250}

Figure \ref{fig:ps250} shows $P(k)$ for 250~\micron\ for the Cyg~X
region, for the original map and with sources removed. The effect of
the beam is dramatic at high $k$, showing that it is important to have
a good estimate of the PSF.  There is also clearly excess power due to
sources and so it is important to remove them carefully.

At small $k$, the power spectrum is dominated by the dust structures;
it decays toward higher $k$, as seen previously in IRIS data at
shorter wavelengths.
The oscillatory effect of $\Gamma(k)$ can be seen clearly at $k >
0.2$~ \arc\ and at higher $k$ the power spectrum meets the
independently-predicted noise level shown.

$P_{\rm{cirrus}}(k)$, obtained by dividing the source-removed,
noise-subtracted power spectrum by $\Gamma(k)$, is also shown in
Figure \ref{fig:ps250}.  It appears to have a power-law form over the
limited dynamic range in $k$ available. We fit only over the range
$0.025\ \rm{arcmin}^{-1} <$ $ k$ $ < 0.2\ \rm{arcmin}^{-1}$,
restricted for the reasons discussed in \S~\ref{sec:resI}.  In
particular, at large $k$ the effect of correcting for the PSF is very
large.  The synthetic PSF is an average for the entire observed field
and so is imperfectly modeled to the precision that would be required
for precise compensation at high $k$.
Convincing evidence for any deviation from a power law for the higher
$k$ range will have to await the higher resolution observations anticipated
with \textit{Herschel} which will probe to beyond $k = 1$.  
At smaller $k$ the observed power spectrum is affected both by
apodization and by the effective filtering of the map-making
procedure, such that the largest scales are not recovered.  For the
relatively small size of these maps and the characteristics of these
BLAST05 observations these effects set in at about the same $k$
\citep{patanchon2008}.
In earlier work with \IRAS , the amplitude $P_0$ is often cited for
$k_0 = 0.01$ \arc.  This fiducial value is below the range of
our observations. Using $k_0 = 0.1$ \arc\ would seem preferable and
probably more relevant to small maps at higher resolution to be made
with \textit{Herschel}.

\begin{figure}
\includegraphics[angle=0,width=\linewidth]{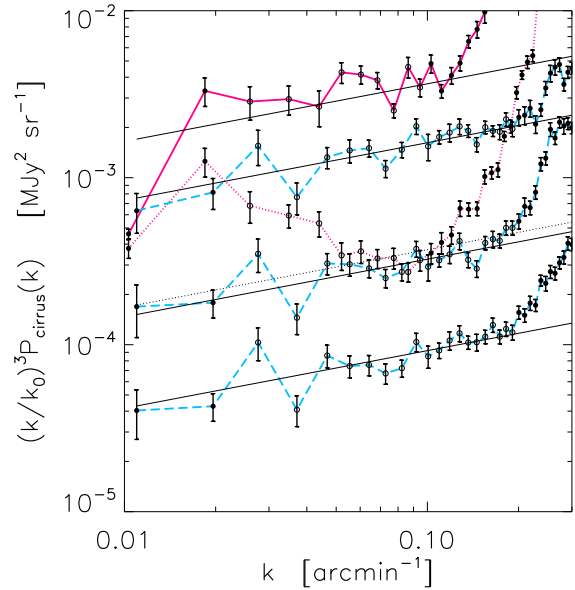}
\caption{$(k/k_0)^3 P_{\rm{cirrus}}(k)$ for the three BLAST bands and
  two IRIS bands, for the Cyg~X field with $k_0\ =\ 0.1$ \arc. The
  exponents are similar for all bands. The amplitude increases with
  decreasing wavelength for the BLAST bands and IRIS 100~\micron\ band
  (solid), and then decreases at 60~\micron\ (dotted). Lines represent
  a power law with common exponent $-2.65$ fit to the data plotted
  with open circles. Data at higher $k$ are not used because of the
  large and uncertain beam correction and eventually the dominance of
  noise (see Figs.~\ref{fig:tf} and \ref{fig:ps250}).}
\label{fig:psblast-cyg}
\end{figure}

Figure \ref{fig:psblast-cyg} shows $(k/k_0)^3 P_{\rm{cirrus}}(k)$ for
all three BLAST bands (250, 350, and 500~\micron) for Cyg~X. We find
$P(0.1\ \rm{arcmin}^{-1}) = $ \camp \ ${\rm MJy}^2\ {\rm sr}^{-1}$ at
250~\micron.  $P_0$ increases with decreasing wavelength for the BLAST
bands and for the IRIS bands (\S~\ref{sec:resI}) remains about the
same at 100~\micron\ and then decreases at 60~\micron.  We obtain an
exponent $\alpha$ equal to \cslope\ at 250~\micron.
This can be compared with the value found above at 100~\micron, $-2.97
\pm 0.23$. The exponents are quite similar for all bands,
except for 60~\micron\ which is discussed further in
\S~\ref{subsec:masscol}.
If the power spectrum were less steep moving into the submillimeter,
this would provide important evidence for a change in the statistical
properties of the emission and the underlying mass column density and
temperature distribution (\S~\ref{subsec:masscol}).  
Definitive searches for such wavelength dependence should benefit from
the higher spatial dynamic range anticipated in observations with
\textit{Herschel}.

\begin{figure}
\includegraphics[angle=0,width=\linewidth]{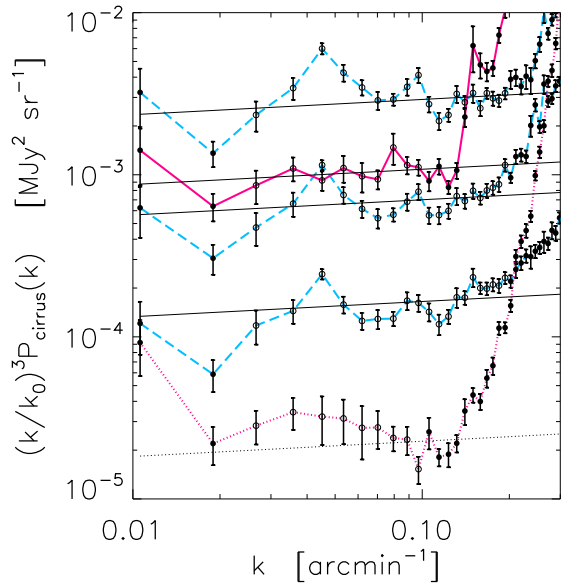}
\caption{Like Fig.~\ref{fig:psblast-cyg}, but for the Aquila region.
  The amplitudes at 100 and 60~\micron\ are relatively lower, which
  indicates that the cirrus in this region is cooler. The common
  exponent used for the fits was $-2.91$.}
\label{fig:psblast-aqu}
\end{figure}

Figure \ref{fig:psblast-aqu} presents the results from our analysis of
the Aquila field.  Overall the same behavior is seen, though the
exponent appears to be slightly steeper (\aslope) at 250~\micron ,
$P_0$ at 250~\micron\ is larger, \aamp\ ${\rm MJy}^2\ {\rm sr}^{-1}$,
and the relative amplitudes at the IRIS bands are much lower.

\subsection{Catalog Depth}\label{sec:depth}

Equation (\ref{sigcir}) can be used to predict the depth reached in a
compact source catalog.  To apply this to the BLAST05 Cyg~X survey,
(where the compact sources are of apparent size $\sim 1.5$\arcmin\ in
the L-R maps in which they are detected and measured), the appropriate
$r$ is $\sim 3.6$. With $D = 1.9\ {\rm m}$ and $P(0.1\ {\rm
  arcmin}^{-1}) = $ \camp\ ${\rm MJy}^2\ {\rm sr}^{-1}$ for the
somewhat dimmer part of the survey field examined here,
$\sigma_{\rm{cirrus}} = 4.2\ {\rm Jy}$ at 250~\micron.  This compares
well to the empirically-estimated 3-$\sigma$ detection threshold of
$\sim 15\ {\rm Jy}$ (Roy et al., in preparation). The effective
noise near bright sources can be somewhat higher because of
contaminating artifacts produced in the L-R deconvolution.

For the large survey of Vela carried out by BLAST in 2006, where the
beam of the 1.8~m mirror was diffraction limited
\citep{netterfield2009}, we have not measured $P_0$ directly, because
the region was not covered by orthogonal scanning and so the cirrus
structure is not well constrained over all scales in the cross-scan
direction \citep{patanchon2008}. Nevertheless, following the scaling
suggested in \S~\ref{sec:dis}, we can estimate $P_0$ from $\langle
I_{100} \rangle \sim 100\ {\rm MJy\ sr}^{-1}$ and $T_{\rm{d}} \sim
18$~K, giving $P(0.1\ {\rm arcmin}^{-1}) = 10^{-3}\ \ {\rm
  MJy}^2\ {\rm sr}^{-1}$ at 250~\micron, from which
$\sigma_{\rm{cirrus}} \sim 0.6\ {\rm Jy}$. However, the sources were
actually extended, with typical apparent sizes of 1\arcmin, so that $r
\sim 2.4$, and then $\sigma_{\rm{cirrus}} \sim 1.6\ {\rm Jy}$. For
comparison, the catalog depth judged from simulations of completeness
was 6~Jy \citep{netterfield2009}. We conclude that the depth is
dominated by the influence of the cirrus noise.

With \textit{Herschel} and assuming the source sizes are compatible
with $r = 1.6$, $\sigma_{\rm{cirrus}}$ in the Cyg~X region should be
closer to 100~mJy at 250~\micron . While a distinct improvement, this
is nevertheless substantial compared to the instrument noise
(r.m.s. $\sim $ 5~mJy) predicted using
HSPOT\footnote{http://www.ipac.caltech.edu/Herschel/hspot.shtml} for
the planned parallel PACS-SPIRE map scanning strategy in the
guaranteed-time key project
HOBYS\footnote{http://starformation-herschel.iap.fr/hobys/} (Herschel
imaging survey of OB Young Stellar objects).
We expect substantial regional variations.  For example, in Aquila,
simply from the larger $P_0$ the depth achieved should be 1.4 times
worse. And this is not even the brightest part of the Galactic Plane
to be surveyed in the open-time key project
Hi-GAL\footnote{https://hi-gal.ifsi-roma.inaf.it/higal} (Herschel
infrared Galactic plane survey).

Extraction of sources with wavelength-dependent and multi-scale
structure will be especially problematical and it will be necessary
both to measure the cirrus properties and to simulate its effects.
For SPIRE maps, in particular those made with the parallel PACS-SPIRE
mode, a corollary is that the high signal-to-noise ratio resulting
from the redundant coverage is well suited to studying the statistical
properties of the bright cirrus.  Of particular interest is how the
power spectrum varies with wavelength, which can be used to determine
physical properties of the diffuse dust.

\section{Wavelength Dependence}\label{sec:dis}

\subsection{ Mass Column Density Distribution} \label{subsec:masscol}

BLAST and \IRAS\ maps record surface brightness $I$ which, in turn,
depends on the dust mass column density $M_{\rm{d}}$:
\begin{equation}
I_\nu = M_{\rm{d}} \kappa_\nu B_\nu(T_{\rm{d}}),
\label{eq:intensity}
\end{equation}
where $\kappa_\nu$ is the dust emissivity, $B_\nu$ is the Planck
function for dust temperature $T_{\rm{d}}$ and it is understood
that the right-hand side is summed over various dust components
which might have different $\kappa$ and $T_d$.

The underlying spatial property being probed is $M_{\rm{d}}$ which
is in turn the projection of the three-dimensional density 
distribution. $M_{\rm{d}}$ is modulated as a function of frequency
by the emissivity and temperature, and can also have
modulations caused by spatial differences in these properties.
The ISM in the Galactic Plane is far from homogeneous, containing
diffuse atomic and ionized regions and shielded molecular regions with
possibly different mixes of dust compositions and sizes, and along any
line of sight there will be a range of interstellar radiation field
intensities, all of which could affect $T_{\rm{d}}$.  The surface
brightness $I_{\nu}$ is particularly sensitive to $T_{\rm{d}}$ for
frequencies near or above the peak in the SED, i.e., for
100~\micron\ and shorter for typical interstellar diffuse dust
temperatures.  Therefore, it is somewhat surprising that the power
spectra exponents that we obtained are so similar.  The potential for
wavelength dependence should be particularly high at 60~\micron\ where
non-equilibrium emission from very small grains (VSGs) starts to
become important \citep{desert1990,lidraine2001} and is even more
directly responsive to the ultraviolet radiation field.
The Cyg~X field analysed definitely has different spatial structures
appearing in the images at the shorter wavelengths and there is some
evidence in Figure~\ref{fig:psblast-cyg} for an effect on the exponent
of the power spectrum.
Furthermore, the exponent  for a particular
region must depend, through the structure of $M_{\rm{d}}$, on the
environments probed along the line of sight and the energy injection
at large scales that is responsible for the apparent turbulence in the
ISM.

In the simple case of homogeneous correlation in the ISM, $I_\nu$
everywhere in the map would scale with frequency according to the
simple relative spectral energy distribution $S_\nu$ of the emitting
dust.  From equation~(\ref{eq:intensity}), $S_{\nu}$ $\propto
\kappa_\nu B_\nu(T_{\rm{d}})$.  Therefore, $P_0$, measuring the same
structure in $M_{\rm{d}}$, would scale simply as $S_{\nu}^2$ (see
eq.~[\ref{eq:pseck}]) while $\sigma_{\rm{cirrus}}$ would scale as
$S_{\nu}$ (see eq.~[\ref{sigcir}]).
This scaling is what is assumed by the prescriptions of
\citet{lagache2000} and \citet{ma-dust}, for example, and appreciated
by \citet{jeong2005}.  However, on making the substitution $P_{100}
\propto \langle I_{100} \rangle^3$, \citet{helou1990} end up assuming
that $P$ scales as $S^3$ and thus $\sigma_{\rm{cirrus}}$ scales as
$S^{1.5}$.  \citet{kiss2001} find some empirical support for the
latter over the \textit{ISO} ISOPHOT wavelength range.

Note that the wavelength dependence of the signal-to-noise ratio of a
point source will vary by the factor $(D/\lambda)^{2.5}$ from
equation~(\ref{sigcir}), and also the ratio of the SED of the source
relative to that of the confusing cirrus.  At submillimeter
wavelengths, the situation is more favorable for detecting cold
sources in warm cirrus than vice versa. Therefore, the completeness
depth of a survey as a function of wavelength will depend on these
factors. Source size and structure can also change with wavelength
\citep{netterfield2009}
but we will not dwell on this quantitatively here.

\subsection{Characteristic Temperature}\label{subsec:temp}

\begin{figure*}
\includegraphics[angle=0,width=0.45\linewidth]{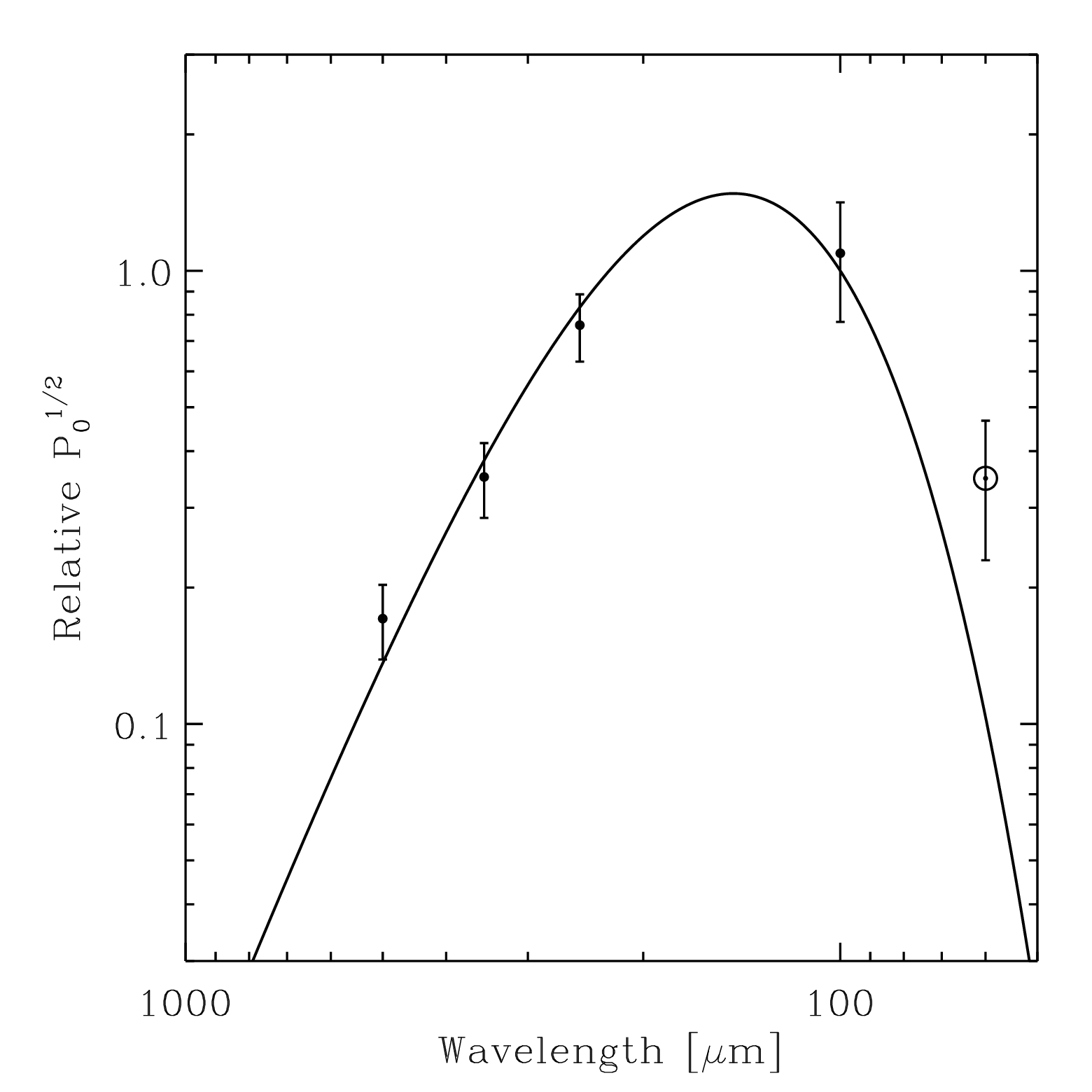}
\includegraphics[angle=0,width=0.45\linewidth]{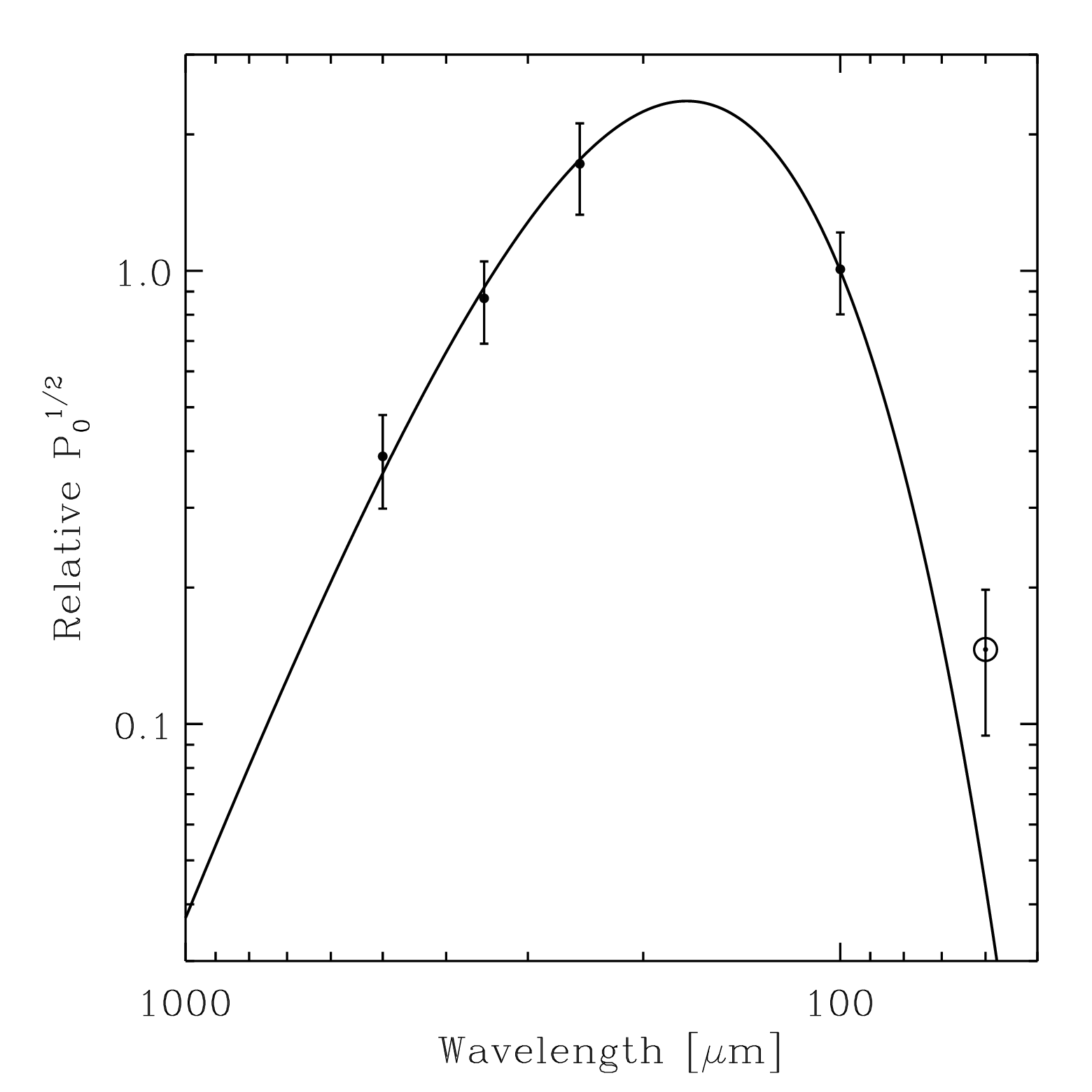}
\caption{Relative SED obtained from the square root of the amplitudes
  of the power spectra fit in Figs.~\ref{fig:psblast-cyg} and
  \ref{fig:psblast-aqu}. Left Cyg~X; right, Aquila. Solid curve is the
  best-fit modified blackbody with $\beta\ =\ \betaval$.  The
  60~\micron\ power spectrum values (open circles) were not used for
  the fits (see text). The best-fit temperatures of Cyg~X and Aquila
  are \cpt\ and \apt~K, respectively. For ease of comparing the two
  fields and results in Figs.~\ref{fig:cor} and \ref{fig:aq-cor}, data
  and curves have been displayed normalized to the value of the
  best-fit modified blackbody at 100~\micron.}
\label{fig:psed}
\end{figure*}

The similarity in the exponents of the power spectra as a function of
wavelength is no guarantee of underlying homogeneous conditions.
Nevertheless, we can assume that there is some characteristic SED for
each region, and use the relative frequency dependence of $P_0^{1/2}$
to recover it. To be consistent with this picture , we evaluated
$P(0.1\ {\rm arcmin}^{-1})$ using a common exponent, which is simply
the average of the exponents found at each wavelength (excluding
60~\micron ).  This relative SED is shown in Figure \ref{fig:psed} for
each region.

Fitting this relative SED with a modified black body
(eq.~[\ref{eq:intensity}]) yields a characteristic dust temperature.
We have assumed $\kappa_{\nu} \propto \nu^\beta$ with $\beta =
\betaval$ \citep{hildebrand1983}.  We did not include the
60~\micron\ data, because the excess there, presumed to be from
non-equilibrium emission, would be inconsistent with a simple
single-temperature modified black body.  For Cyg~X, the increase at
long wavelengths with respect to the fit is probably the result of the
range of temperatures known to be present in different parts of this
region (Roy et al., in preparation).  Therefore, even in this case we
kept $\beta = 2$, although empirically $\beta = 1.5$ would produce a
better fit.

For Cyg~X, the best fit temperature is \cpt\ K, which appears to be
slightly warmer than the local high latitude diffuse medium dust
temperature of 17.5~K \citep{boulanger1996}.  Finding warmer dust in
Cyg~X is not particularly surprising, because this region, which lies
along the local spiral feature ($l = 80 $\degree ), is one of active
star formation, with an OB association and numerous HII regions.  In
the 21-cm radio continuum this region includes areas of strong diffuse
thermal emission.  To account for the higher equilibrium temperature
the ambient radiation field absorbed in Cyg~X would have to be about
$(19.9/17.5)^6 = 2.1$ times higher than locally.

By contrast, for Aquila the characteristic temperature derived is
somewhat lower, about \apt\ K.  This line of sight passes through the
inner Galaxy ($l = 45$\degree ), with significant molecular clouds but
much less star formation (Rivera-Ingraham et al., in preparation).

\begin{figure*}
\includegraphics[angle=0,width=0.45\linewidth]{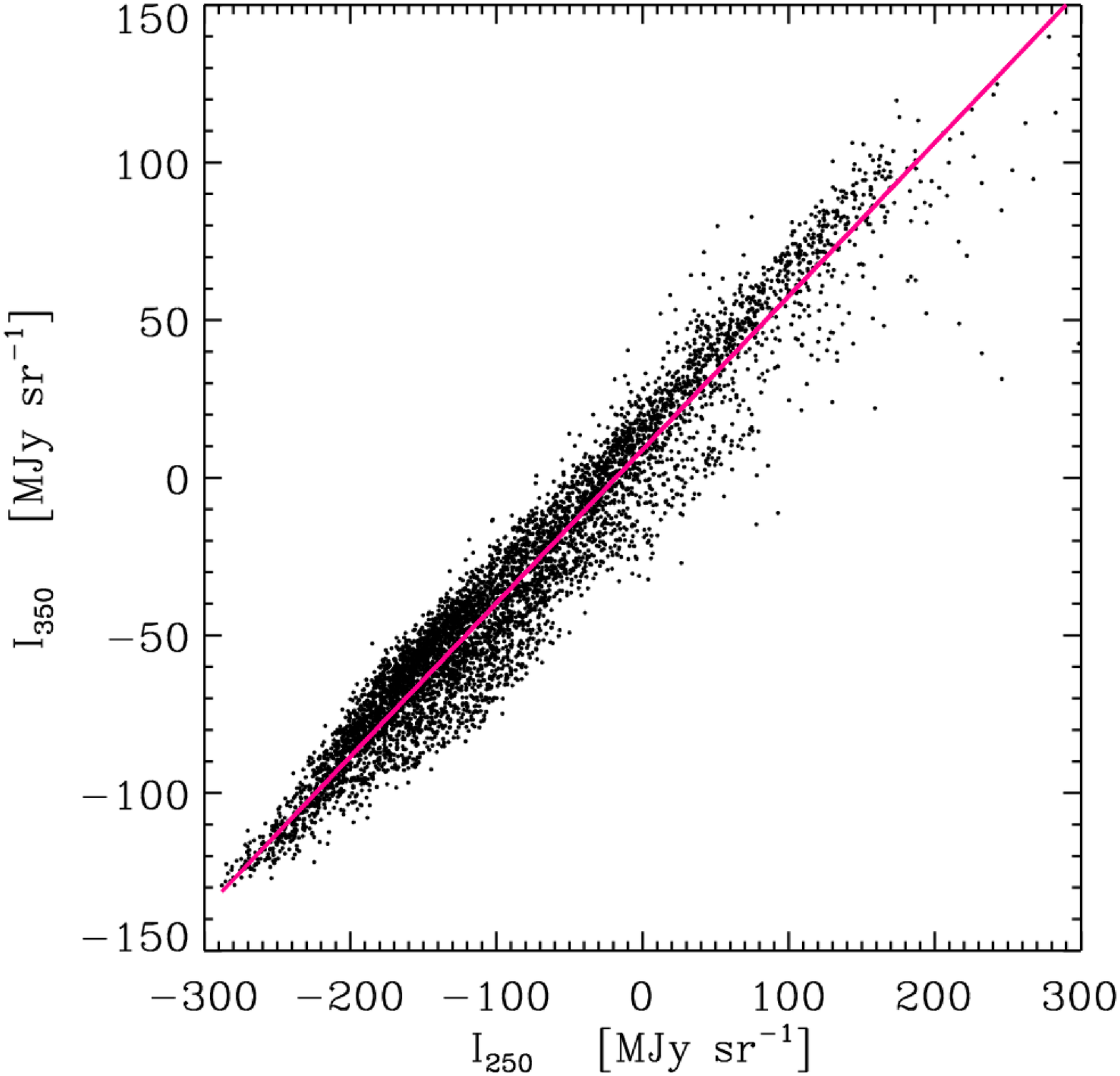}
\includegraphics[angle=0,width=0.45\linewidth]{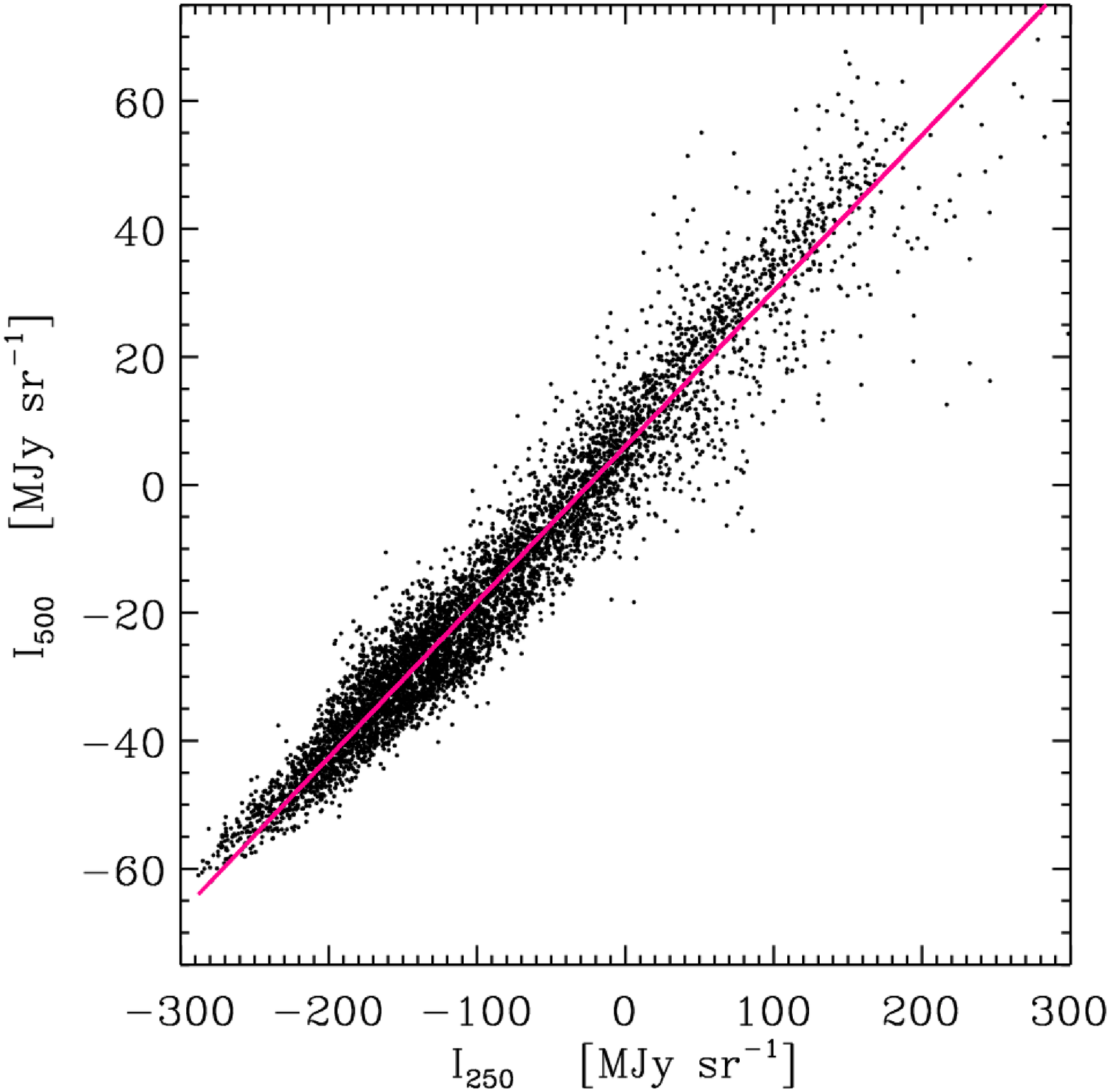}
\includegraphics[angle=0,width=0.45\linewidth]{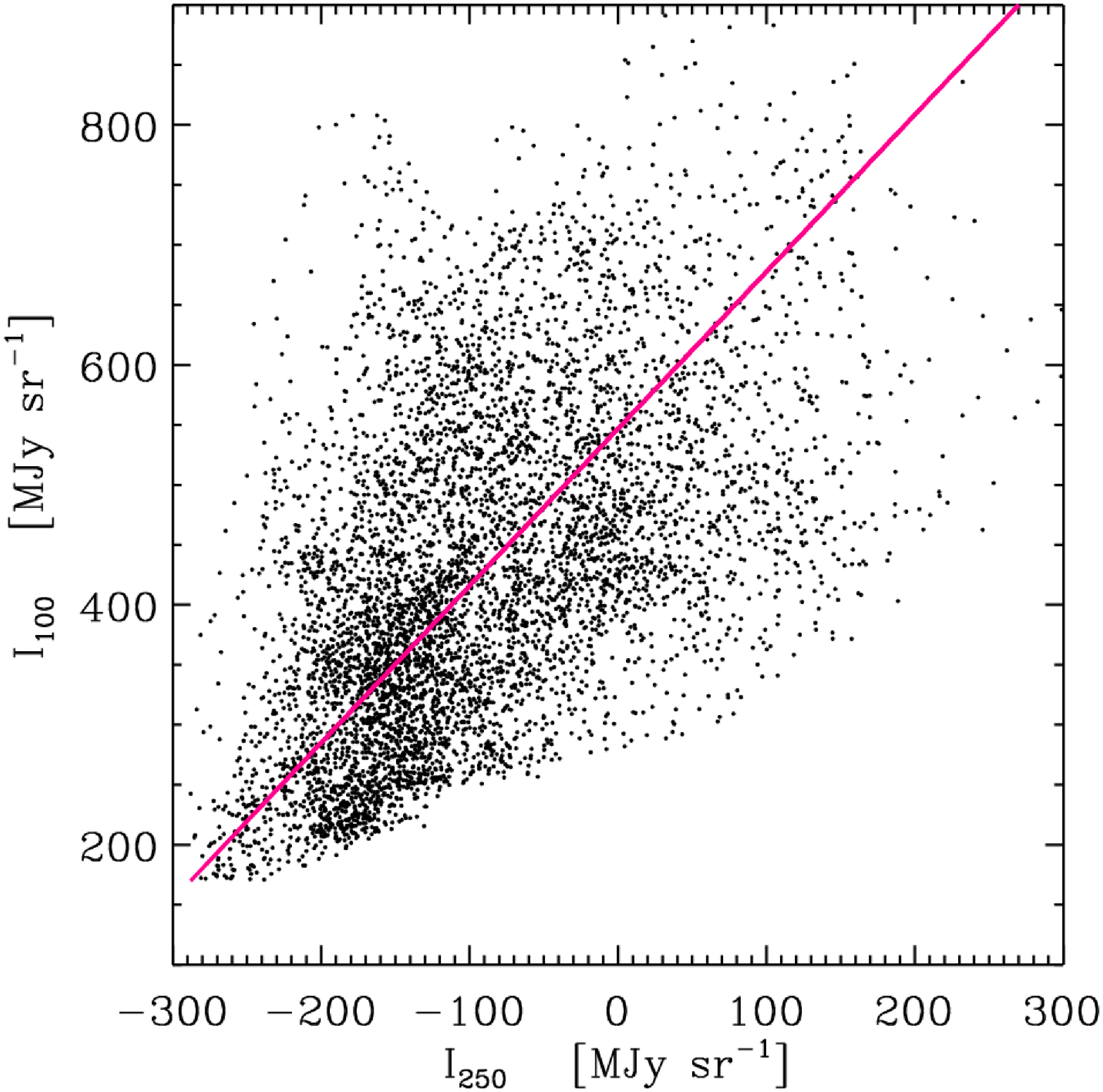}
\includegraphics[angle=0,width=0.45\linewidth]{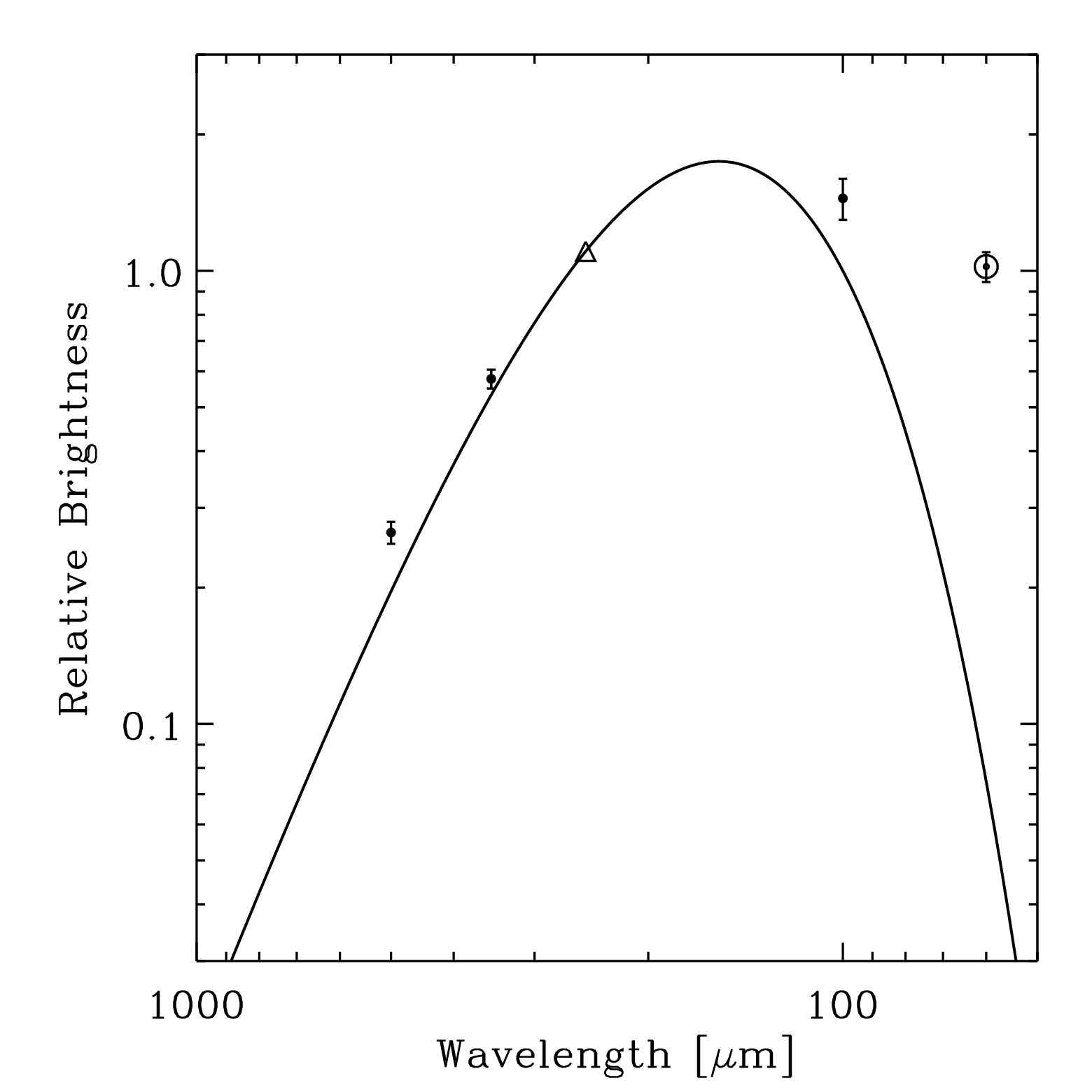}
\caption{Pixel-pixel correlation of 350, 500, and 100~\micron\ maps
  with respect to the 250~\micron\ map, with the best-fit correlation
  line plotted. The lower right panel shows a relative SED obtained
  from the slopes of these correlations. With $\beta=\betaval$, the
  best-fit temperature is \cct~K.  The normalization here is the same
  as in Fig.~\ref{fig:psed}. The value at 250~\micron, indicated by
  the triangle in this normalization, is not used explicitly for the
  fit. The open circle (from the 60~\micron\ slope) was also not
  used.}
\label{fig:cor}
\end{figure*}

\begin{figure*}
\includegraphics[angle=0,width=0.45\linewidth]{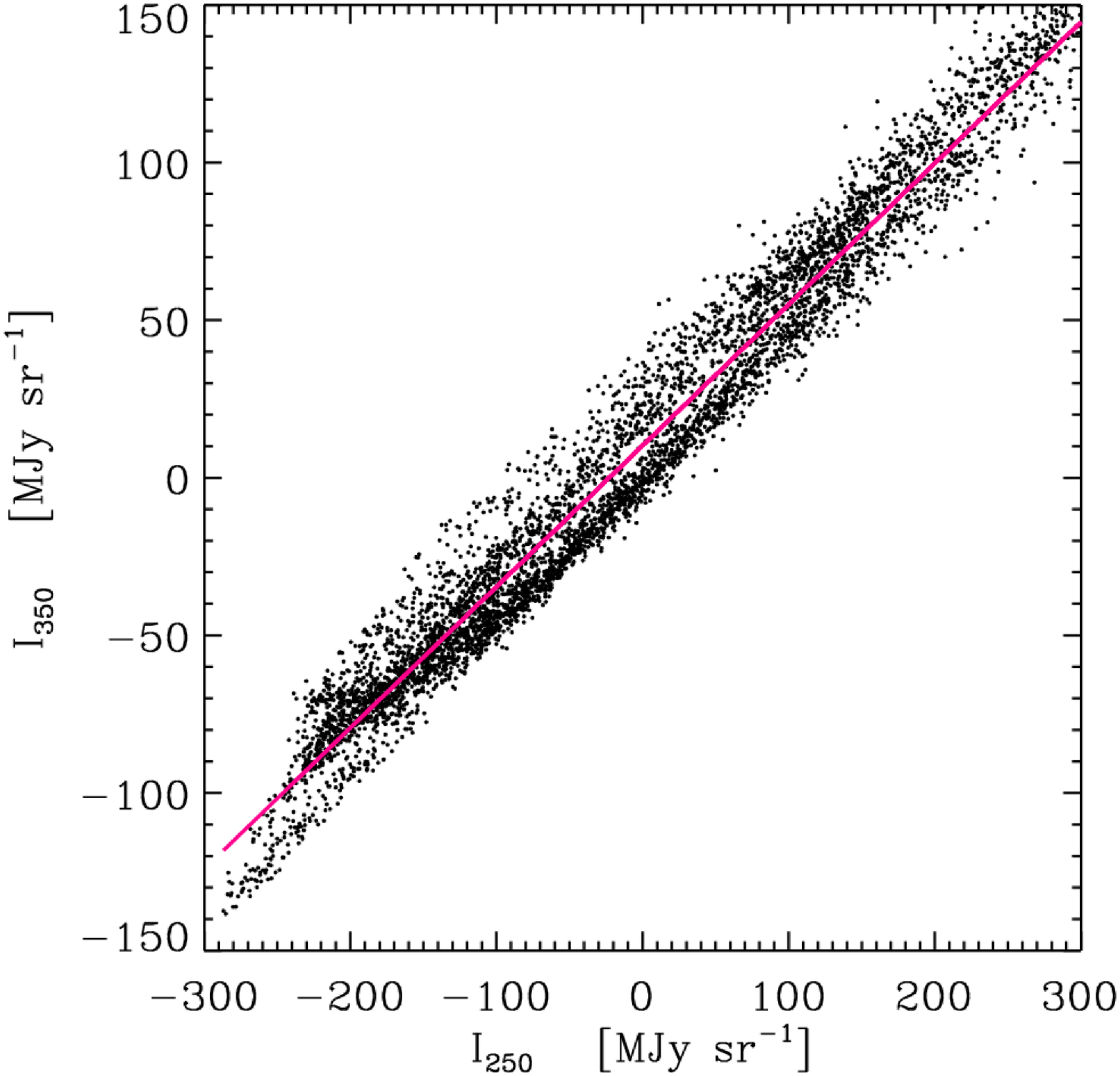}
\includegraphics[angle=0,width=0.45\linewidth]{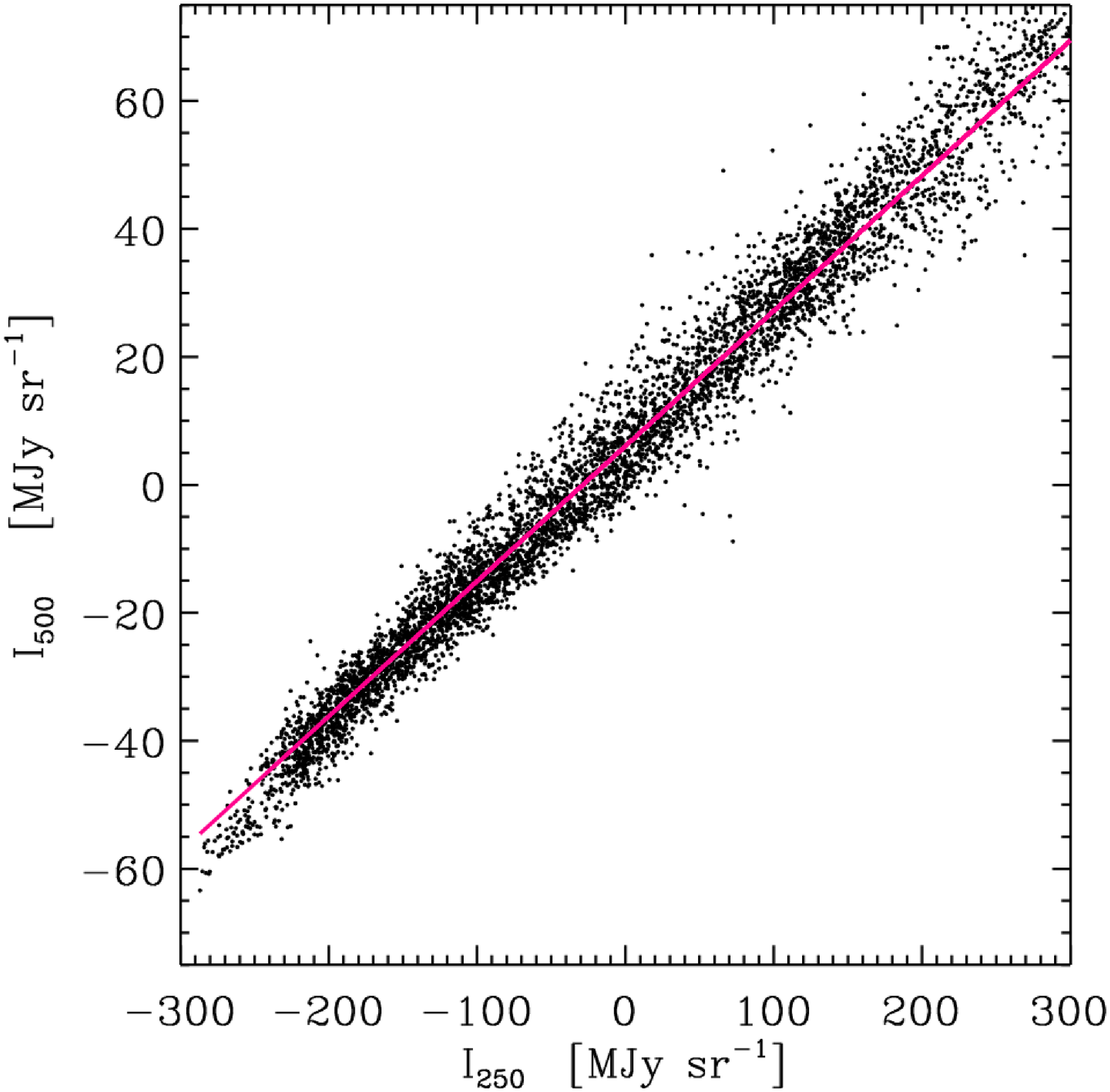}
\includegraphics[angle=0,width=0.45\linewidth]{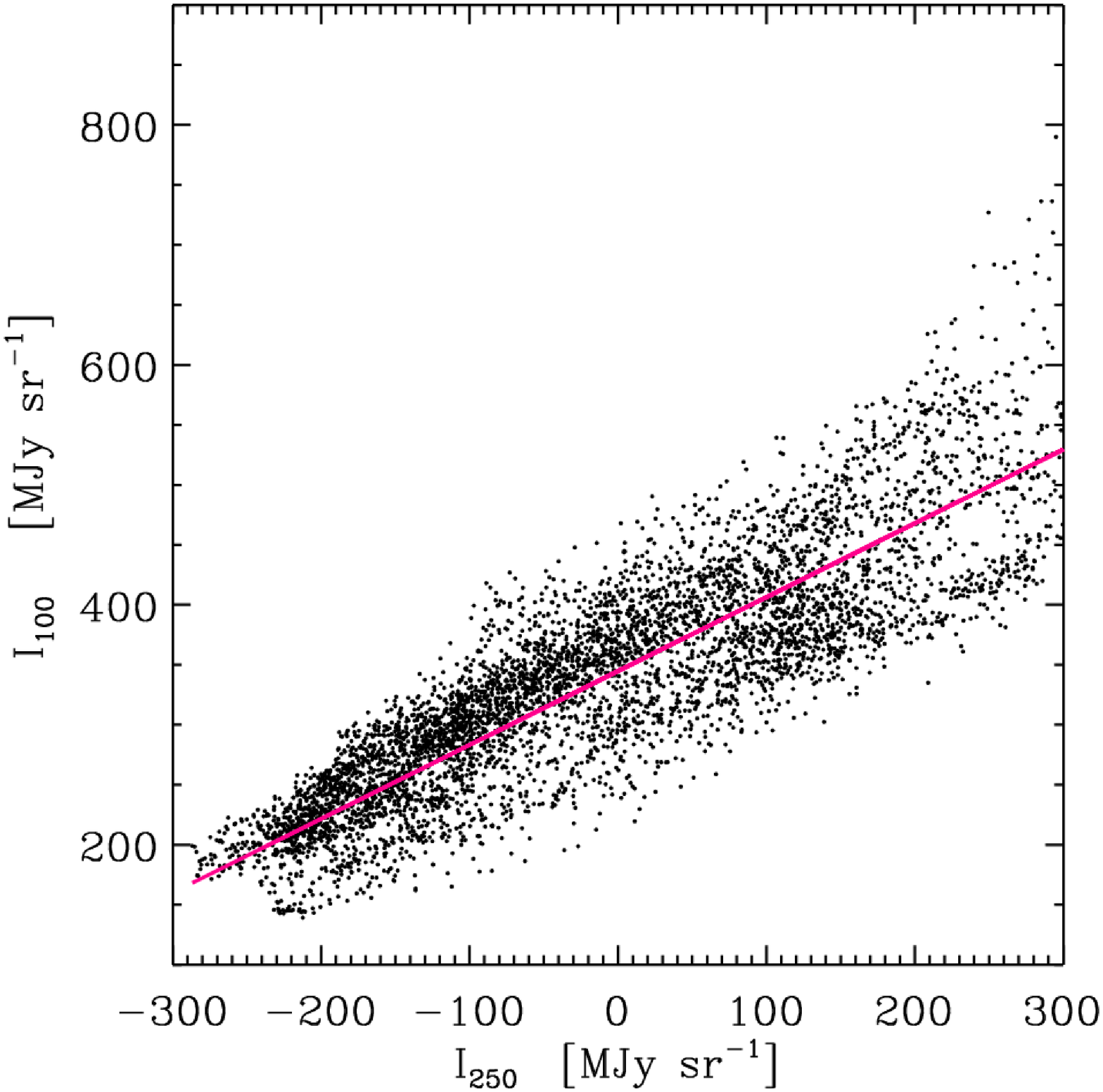}
\includegraphics[angle=0,width=0.45\linewidth]{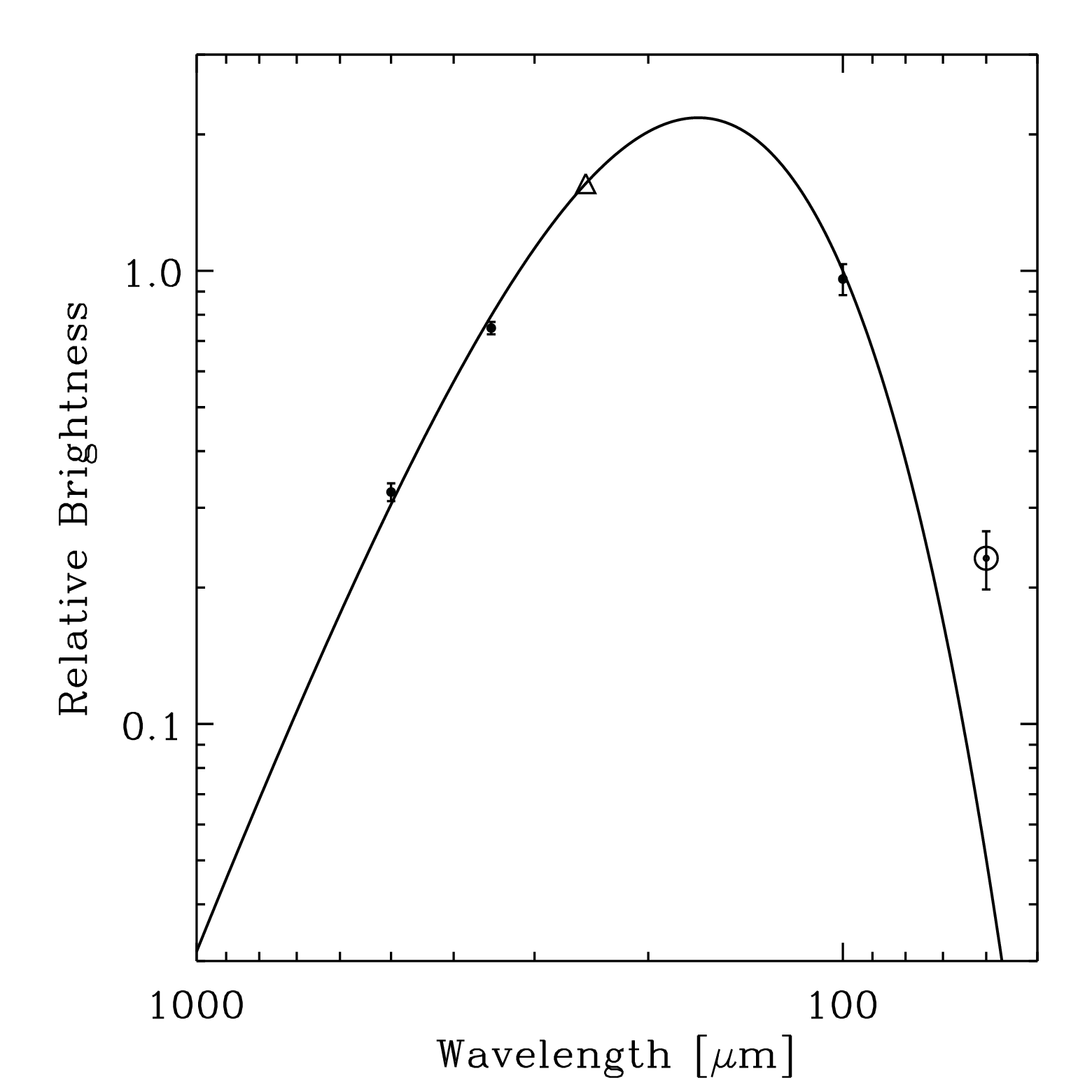}
\caption{Like Fig.~\ref{fig:cor}, but for the Aquila region. The slope
  of the 100~\micron\ correlation is much shallower than in the Cyg~X
  field, and the best-fit temperature is \act\ K.}
\label{fig:aq-cor}
\end{figure*}

As we have noted, under uniform conditions, maps at different
wavelengths will have to be highly correlated in their spatial
structures, and so they will have the same power spectrum scaled by
$S^2$. But the fact that the relative SED derived from $P_0^{1/2}$
appears reasonable is no guarantee that the maps are simply scaled
versions of one another (since phases could be different). However,
this can be checked directly.  Inspection of the BLAST maps shows that
they are remarkably similar in all three bands.  This is quantified
through the correlations with respect to the 250~\micron\ map shown in
Figure \ref{fig:cor} for Cyg~X.  The linear fits shown are the
ordinary least squares bisector solution \citep{Isobe} obtained with
the IDL routine SIXLIN.  Note that the correlation of BLAST
250~\micron\ with IRIS 100~\micron\ is not as high as it is between
BLAST bands. This is expected, because of the above-mentioned greater
sensitivity of 100~\micron\ emission to $T_{\rm{d}}$.
Differences with respect to the submillimeter images become more
apparent at 60~\micron, where in addition the power spectrum is found
to be somewhat steeper (Fig.~\ref{fig:psblast-cyg}).
Figure~\ref{fig:aq-cor} shows the even better correlations for the
Aquila region. A clue to understanding the good correlation is, of
course, the very detailed similarity of the power spectra at the
different wavelengths.

As a consistency check, the slopes of the correlations can be used to
construct another relative SED.  As seen in Figures~\ref{fig:cor} and
\ref{fig:aq-cor}, these look very similar to the respective SEDs based
on the power spectra (Fig.~\ref{fig:psed}) and the best-fit
temperatures \cct~K and \act~K are very similar to those found above.

We therefore have some evidence that the appropriate scaling of
$\sigma_{\rm{cirrus}}$ is as $S$, not $S^{1.5}$.  However, this is not
immediately generally useful, unless there is some independent
evidence about the frequency dependence of $S$ (or the effective
$T_{\rm{d}}$ and $\beta$), since this can clearly vary significantly
from region to region.  The best approach is to measure $P_0$ directly
at each wavelength.

\section{Conclusion}\label{sec:con}

In these fields in the Galactic Plane, the exponent of the power
spectrum of the 100~\micron\ IRIS map is close to $-3$, within the
dispersion seen by \citet{ma-dust} despite the average brightness
$\langle I_{100} \rangle$ being beyond the range studied by these
authors.
On the other hand, the amplitudes of the 100~\micron\ power spectra
estimated for these bright $\sim 2$\degree $\times$ $2$\degree\ fields
are significantly below what would be extrapolated from the trend with
$\langle I_{100} \rangle$ found by \citet{ma-dust}.  Therefore, 
particularly in such bright star forming regions, it
is recommended that the power spectrum be computed directly.
The power spectra derived from the BLAST observations are also well
fit by power laws, with similar exponents.
The frequency dependence of the amplitude of the power spectrum can be
described by the square of an SED which is a simple modified black
body function with a reasonable characteristic temperature.
This is confirmed by direct correlations between the maps at different
wavelengths.
However, this characteristic temperature does appear to change in
different Galactic environments, and unless its value is known
independently the power spectra and/or map correlations need to be
evaluated for all wavelengths.

Cirrus noise will be important in many planned multi-wavelength
Galactic Plane, high latitude, and extragalactic surveys carried
out with \textit{Herschel}.
Our results provide important empirical support for a proposed
prescription for the wavelength dependence of the cirrus noise, which
incorporates a factor which varies directly as the SED.  In practice
the noise will be best evaluated, most free of assumptions, simply by
measuring $P_0^{1/2}$ at each wavelength and using
equation~(\ref{sigcir}).

\acknowledgments The BLAST collaboration acknowledges the support of
NASA through grant numbers NAG5-12785, NAG5-13301 and NNGO-6GI11G, the
Canadian Space Agency (CSA), the UK Particle Physics \& Astronomy
Research Council (PPARC), and Canada's Natural Sciences and
Engineering Research Council (NSERC). We would also like to thank the
Columbia Scientific Balloon Facility (CSBF) staff for their
outstanding work.

\bibliographystyle{apj}
\bibliography{msbib}
\end{document}